\documentclass{elsart}

\usepackage{epsfig}
\usepackage{amssymb}

\newcommand{\ie}{{\it i.e. \/}}
\newcommand{\eg}{{\it e.g. \/}}
\newcommand{\cf}{{\it cf. \/}}

\newcommand{\eff}{\mathrm{eff}}

\newcommand{\sect}[1]{Sect.~\ref{#1}}
\newcommand{\fig}[1]{Fig.~\ref{#1}}
\newcommand{\emb}[1]{(\ref{#1})}
\newcommand{\eq}[1]{Eq.~(\ref{#1})}

\newcommand{\tens}[1]{{\stackrel{\leftrightarrow}{#1}}}

\newcommand{\lb}{\left}
\newcommand{\rb}{\right}

\def\dpar#1#2{\frac{\partial#1}{\partial#2}}
\def\ddpar#1#2{\frac{\partial^2#1}{{\partial#2}^2}}
\def\drpar#1#2#3{\frac{\partial^2#1}{\partial#2\partial#3}}

\def\bnabla{\mbox{\boldmath $\nabla $}}

\journal{Computer Physics Communications}

\begin{document}

\begin{frontmatter}

\title{
Poisson -- Boltzmann Brownian Dynamics \\
of \\
Charged Colloids in Suspension
}

\author[UK]{J. Dobnikar}
\author[UM]{D. Halo\v{z}an}
\author[UM]{M. Brumen}
\author[UK]{H.-H. von Gr\"unberg}
\author[IFF]{R. Rzehak\corauthref{cor}}
\ead{r.rzehak@fz-juelich.de}
\corauth[cor]{Corresponding author.}
\address[UK]{%
  Fakult\"at f\"ur Physik
  Universit\"at Konstanz
  D-78457 Konstanz}
\address[UM]{%
  Pedagogical Faculty,
  University of Maribor, Koro\v{s}ka c. 160, SI-2000 Maribor}
\address[IFF]{%
  Institut f\"ur Festk\"orperforschung,
  Forschungszentrum J\"ulich,
  D-52425 J\"ulich}

\begin{abstract}
We describe a method to simulate the dynamics of
charged colloidal particles suspended in 
a liquid containing dissociated ions and salt ions.
Regimes of prime current interest are those of
large volume fraction of colloids,
highly charged particles and 
low salt concentrations. 
A description which is tractable under these conditions is obtained by 
treating the small dissociated and salt ions as continuous fields, 
while keeping the colloidal macroions as discrete particles.
For each spatial configuration of the macroions, 
the electrostatic potential arising from all charges in the system
is determined by solving the nonlinear Poisson--Boltzmann equation. 
From the electrostatic potential, 
the forces acting on the macroions are calculated
and used in a Brownian dynamics simulation 
to obtain the motion of the latter.
The method is validated by comparison to known results
in a parameter regime where the effective interaction between the macroions
is of a pairwise Yukawa form.
\end{abstract}

\begin{keyword}
Poisson-Boltzmann equation \sep Brownian dynamics \sep charge-stabilized colloids
\PACS 82.70.Dd
\sep  02.70.Ns
\sep  41.20.Cv
\sep  64.70.Dv
 
\end{keyword}

\end{frontmatter}

\section{Introduction}
\label{sec-intro}

Systems of charged particles of mesoscopic size
suspended in a liquid
are abundant in fields ranging from biology to chemical engineering
\cite{Evans:CD-94}.
Next to colloids, examples include
proteins, polyelectrolytes, micelles etc. 
\cite{Safran:PCSF87,Daoud:SMP-99}.
However, despite great interest 
from both fundamental and applied points of view,
the phase behavior and transport properties of such systems
are still only partly understood.

The mesoscopic particles become charged because
in a polar solvent like water
small ions dissociate off of their surface.
In addition to these dissociated ions
the solution often also contains salt ions.
In the following, no distinction will be made 
between dissociated ions 
and salt ions carrying the same charge.
All the small ions
are of molecular size and
carry one or at most a few elementary charges.
The mesoscopic particles, in contrast,
have a much higher charge on them, hence,
they are commonly referred to as macroions.
Due to the electrostatic interaction,
the oppositely charged counter-ions in the solution
concentrate around the macroions
while the like charged co-ions are depleted.
Entropy opposes this tendency.
As a result, an inhomogeneous distribution of small ions is established
and part of the bare charge on the macroions is screened.
The charges on the surface of the macroion
together with the screening mobile charges in the solution 
are commonly referred to as the diffuse electric double layer 
\cite{Carnie:ACP56-84-141,Attard:ACP92-96-1}.

When the double layers of two or more macroions overlap,
the latter experience an effective force.
Attempts to predict the macroscopic suspension properties
mostly start from an assumed form of this interaction 
which is taken as pairwise additive 
\cite{Belloni:JPHC12-00-R549,Likos:PR348-01-267}.
The validity of this assumption depends on
the range of the double layer interaction which varies with 
the salt concentration and the charge on the macroions.
For high salt concentration,
the interaction range is small compared to 
the average distance between the macroions
so that no more than two double layers overlap simultaneously.
Thus, the double layer interaction is indeed pairwise additive
and according to DLVO theory \cite{Verwey:TSLC-48,Bell:JCIS33-70-335}
described by a repulsive Yukawa potential.
The effect of the macroionic charge in this case can be accounted for by 
renormalizing charge and size of the macroions 
\cite{Alexander:JCP80-84-5776,Belloni:CSA140-98-227,Levin:RPP65-02-1577}.
At low salt concentrations,
where the interaction range is 
several times the average distance between the macroions,
it is very likely that 
the double layers of more than two macroions overlap at the same time.
Hence, many-body effects become important and 
the description of suspension properties by a pair-potential
becomes questionable. 
An explicit form for the additional many-body forces that arise, however,
is not known at present.

In the broad range of parameters where 
the form of the effective interaction is not known,
a theoretical description must include the small ions explicitly.
In the primitive model \cite{Vlachy:ARPC50-99-145}
both macroions and small ions are treated as 
spherical particles with different size and charge
suspended in a continuous polarizable solvent. 
The feasibility of simulations using this model is limited by
the number of particles involved:
For highly charged macroions, many small ions are necessary to 
satisfy the constraint of overall charge neutrality
and significant amounts of salt present in the solution
further increase the number of particles.
The most efficient methods currently allow simulation of 
only 80 mesoscopic particles
each carrying no more than 60 elementary charges
and with no salt added to the solution \cite{Lobaskin:JCP111-99-4300}.
In practice, 
the macroions carry up to several thousand elementary charges
and salt concentrations between micromoles and millimoles
per liter are encountered 
while the volume fraction of the macroions is at most a few percent.
Clearly, under these conditions
primitive model calculations become intractable 
for reasonably large systems
and a different approach is called for. 

A reduced description of the system is obtained by exploiting the fact 
that the small ions, having a much smaller mass, 
move on a much faster time scale than the macroions. 
This permits an adiabatic approximation, where 
partial equilibrium of the small ions for 
the instantaneous configuration of the macroions is assumed.
Consequently, the small ions are described by a continuous density
which minimizes an appropriate thermodynamic potential 
\cite{Reiner:JSCFT86-90-3901,Lowen:JCP98-93-3275,Netz:EPJE1-00-203,Deserno:ESMB-02-27}. 
Using this density in the Poisson equation for the electrostatic potential
results in a closed relation for the latter. 
When correlations between the small ions are neglected, 
this relation becomes the nonlinear Poisson-Boltzmann ({\bf PB}) equation 
\cite{Quarrie:SM-73,Israelachvili:ISF-85}. 
This mean--field description of the electrostatic effects 
has been shown to be a good approximation for
the case of monovalent small ions and weak coupling between the small
ions by comparison to a cell model Monte-Carlo simulation
\cite{Groot:JCP95-91-9191}.
A more complete description of the electrostatics can be obtained by
using more sophisticated free energy functionals 
\cite{Lowen:JCP98-93-3275,Lowen:EL23-93-673}.

Even on the PB level of description,
the physics of colloidal suspensions presents a challenging problem
so that further approximations are commonly invoked.
At high volume fraction,
the many-body problem posed by the suspension
is reduced to an approximate one-body problem
by using cell models \cite{Alexander:JCP80-84-5776,Deserno:ESMB-02-27}.
However, the full nonlinear PB equation can be solved analytically 
only in very few cases \cite{Andelman:HBP-95-12}.
Even for the extremely simple geometry of a spherical cell,
no analytic solution is available.
Therefore, its linearized version has often been used as a substitute
in deducing suspension properties 
\cite{Deserno:PRE66-02-011401,Tamashiro:NN1-02}.
By the combination of modern high-speed supercomputers
and efficient numerical algorithms 
a treatment of the full nonlinear many-body problem 
is now within reach.

In this paper, a method combining 
a PB field description of the small ions
with a Brownian dynamics simulation of the macroions
\cite{Fushiki:JCP97-92-6700,Gilson:JCoC16-95-1081} is discussed.
For each spatial configuration of the macroions, 
the electrostatic potential due to all charges in the system
is determined by solving the PB equation 
with boundary conditions determined by 
the positions of the macroions and the charges on them. 
The force acting on each of the macroions 
is then calculated by integrating the stress tensor 
over a surface enclosing the respective macroion.
These forces are finally used in a Brownian dynamics simulation 
to obtain the motion of the macroions.
This method allows to calculate 
structural and thermodynamic properties of
charge-stabilized colloidal suspensions
with full account of the many-body interaction between the colloids
mediated by the screening ionic fluid around them.

To validate the method we here focus on a parameter range where 
the assumption of effectively pairwise additive interactions 
is expected to hold true.
Specifically, we determine the effective pair-force for 
fixed arrangements of the macroions
from the numerical solution of the PB equation and
compare the results to linearized DLVO theory and
predictions based on a cell model.
This provides a stringent test of 
the crucial step in our simulation method,
namely the calculation of the true forces acting on the macroions
which serve as input for the Brownian dynamics part.
The latter is then used to locate the melting point
which for Yukawa systems is known from the classic work of
Robbins, Kremer and Grest \cite{Robbins:JCP88-88-3286}.
The excellent agreement for all of our comparisons
in the regime of pairwise additive effective interactions
confirms our method and implementation.
Outside this parameter regime, many-body effects become important
which are investigated elsewhere \cite{Dobnikar:NN1-02,Dobnikar:NN2-02}.

The paper is organized as follows:
In \sect{sec-system} we 
define the system considered, 
collect the equations of motion,
and discuss the different parameter regimes.
The discretization of the PB equation and the method of solution 
are described in \sect{sec-pbe} while
the force calculation and the Brownian dynamics algorithm 
are summarized in \sect{sec-force} and \sect{sec-md}, respectively.
\sect{sec-test} presents results validating the method
and illustrating its range of applicability.
In \sect{sec-conc} we finally discuss the results and 
point out directions for future development.

\section{System and Equations}
\label{sec-system}

\begin{figure}[h]
  \begin{center}
    \epsfig{file=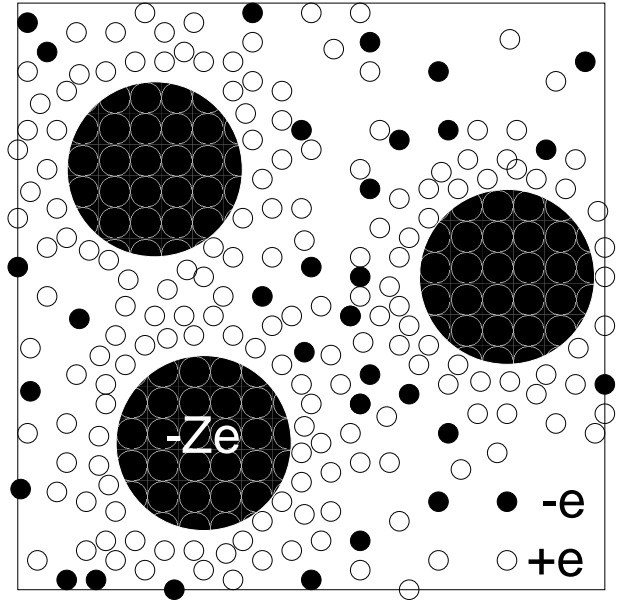, height=8cm}
  \end{center}
  \caption{\label{fig_system}
    Illustration of a suspension with $N=3$ macroions of charge $-Z e$
    suspended in a liquid of small ions with charges $\pm e$. 
  }
\end{figure}
We consider 
$N$ spherical macroions (\cf \fig{fig_system})
in a cubic box of size $L$
which is continued periodically in all three space directions
to minimize finite size effects.
The macroions have a radius $a$ and
and carry a charge $- Z e$
which is distributed evenly over their surface.
Two kinds of small ions with charge $\pm e$ 
may enter the system from a reservoir with ion pair concentration $c_{s}$.
Thus, the number of small ions is not fixed
but varies such that the chemical potential is constant everywhere in
the suspension.
In the absence of charges on the macroions ($Z = 0$),
the average concentration of each species of small ions is $c_{s}$
while otherwise their concentrations adjust so that
charge neutrality is obeyed on average. 
Due to the Donnan effect,
the total number concentration of small ions of any kind
then is smaller than $2 c_{s}$ \cite{Deserno:PRE66-02-011401}.
The presence of other, uncharged molecules in the solution
is accounted for by a dielectric constant $\epsilon$.
Finally, the whole system is coupled to a heat bath
of temperature $T$.

In addition to the macroion size $a$,
there are several other important length scales
which serve to distinguish different parameter regimes.
Denoting the thermal energy as $\beta = 1 / k_{\rm B} T$,
the Bjerrum length, 
\begin{equation}
\label{lambda_def}
\lambda_{B} = \beta\frac{e^2}{\epsilon}
\;,
\end{equation}
gives the length below which
the Coulomb interaction between two small ions
dominates the thermal energy.
Its value gives a measure of the importance of 
correlations between the small ions
and, hence, the quality of the PB description of the electrostatic problem.
In case of monovalent small ions 
primitive model simulations and PB calculations for a spherical cell model 
have been compared in Ref.~\cite{Groot:JCP95-91-9191}. 
The deviations between both decrease with 
the value of the parameter $\lambda_{B}/a$
and are already tiny at  $\lambda_{B} \le 0.03 a$. 
For an aqueous solvent at room temperature, this corresponds to $a \ge 24$nm.  
Throughout our calculations we use 
an even smaller value $\lambda_{B} / a = 0.012$
which corresponds to a particle size of $a=60nm$ 
as in recent experiments \cite{Yamanaka:PRL80-98-5906}.
Hence, we are assured that PB theory furnishes an excellent
description of the systems under investigation here.

The width of the double layer
and hence the range of the effective interaction between the macroions
is given by the Debye screening length
$\kappa^{-1}$, where 
\begin{equation}
\label{kappa_def}
\kappa^{2} = 8 \pi \lambda_{B} c_{s}
\;,
\end{equation}
which emerges from a dimensional analysis of 
the PB equation.
The volume available to each of the macroions, finally, yields 
a measure for the mean distance $d_m$ between two macroions as
\begin{equation}
\label{d_def}
d_m = N^{-1/3} L = \lb( \frac{4 \pi}{3 \eta} \rb)^{1/3} a
\;,
\end{equation}
where the volume fraction $\eta=N \frac{4 \pi}{3} \frac{a^3}{L^3}$ 
has been introduced.
For $\kappa^{-1} \ll d_m$ the interactions between the macroions are 
expected to be effectively pairwise additive.
A sufficient condition that the potential $\psi$ remains small everywhere
is that in addition $Z\lambda_{B} \ll a$. This may be used as
a criterion ensuring that linearization of the PB equation will be valid.

\subsection{Electrostatic problem}
\label{ssec-elstat}

As discussed in the introduction, 
the Poisson-Boltzmann ({\bf PB}) equation is obtained by combining
the Poisson equation for the electrostatic potential with
the charge densities of the two species of small ions
resulting from minimization of 
a mean-field free energy functional $\mathcal{F}$.
The latter is composed of two terms
representing the energy of all charges, fixed and mobile,
in the electrostatic potential generated by them and
the entropy for an ideal gas of small ions 
in contact with a particle reservoir
\begin{eqnarray}
\label{free-energy}
\beta \mathcal{F}
=
\int_{G}                                                &\quad&
  \frac{1}{2} \, \psi \, 
  \lb( n_{fixed} + n_{+} - n_{-} \rb) \\
                                                        &+& 
  n_{+} \lb( \log\!\lb(\frac{n_{+}}{c_s}\rb) - 1 \rb)
+ n_{-} \lb( \log\!\lb(\frac{n_{-}}{c_s}\rb) - 1 \rb) 
\, d^3r                                                 \nonumber
\,.
\end{eqnarray}
Here, 
$\pm e n_{\pm}(\vec{r})$ are the number densities of 
the two species of small ions,
$\psi(\vec{r}) = \beta e \phi(\vec{r})$ is the scaled potential, 
and the integration volume $G$ is the region outside the macroions.
Minimization subject to the constraint of 
electro-neutrality gives
\begin{equation}
\label{n_eq}
n_{\pm} = c_{s} \exp\!\lb( \mp \psi \rb)
\;.
\end{equation}
Together with the Poisson equation this leads to the PB equation
which holds in the region $G$ outside the macroions, \ie
\begin{equation}
\label{PBE}
\bnabla^{2} \psi = \kappa^{2} \sinh\!\lb(\psi\rb)
\qquad \mbox{for  $\vec{r} \in G$}
\;,
\end{equation}
where $\kappa^{2}$ has been defined in equation \ref{kappa_def}. 
Inside the macroions, where there are no mobile charges, 
the electrostatic potential in principle is governed by 
Poisson's equation and both regions are connected by 
a jump condition accounting for the charge distribution 
on the macroion surfaces.  
However, in aqueous solution, 
the dielectric constant inside the macroions 
is typically much smaller than that of the solvent outside 
and the matching condition between inside and outside 
simplifies to a von Neumann condition for the PB equation.
Denoting the surface of the $p$-th macroion by $\partial G_p$ and its
normal vector pointing into the solvent by $\hat{\vec{n}}_p$, this
boundary condition
is expressed as 
\begin{equation}
\label{BC_coll}
\hat{\vec{n}}_p \cdot \bnabla \psi = \frac{- Z\lambda_B}{a^{2}}
\qquad \mbox{for  $\vec{r} \in \partial G_p$}
\;.
\end{equation}
Together with periodic boundary conditions on the sides of the box 
this furnishes a complete description of the electrostatic part of the problem
from which other quantities can be derived.

\subsection{Force Calculation}
\label{ssec-force}

The forces on the particles 
needed for the Brownian dynamics simulation
are calculated from the stress tensor,
which is a sum of two parts, 
an osmotic and an electrostatic one.
The osmotic part of the stress tensor $\Pi$ is proportional to 
the difference of the total number density of small ions 
from the reservoir concentration, \ie
\begin{eqnarray}
\beta\tens{\Pi}
=
\lb( n_{+} + n_{-} - 2c_{s} \rb) \tens{1}
=
2 \, c_{s} 
\lb( \cosh\!\lb( \psi\rb) - 1 \rb) \tens{1}
\;.
\end{eqnarray}
The electrostatic part of the stress tensor is
\begin{eqnarray}
\beta\tens{T}^{el}=
\frac{1}{8\pi\lambda_B}
\lb((\bnabla \psi)^{2}\tens{1} - 2\bnabla \psi \otimes \bnabla \psi\rb)
\;.
\end{eqnarray}
Together we have
\begin{eqnarray}
\beta\tens{T}                        &=&
\beta\lb( \Pi + \tens{T}^{el} \rb)   \\
                                     &=&
\frac{1}{8 \pi \lambda_B}
\lb(   \lb(   2 \kappa^{2} \lb( \cosh(\psi) - 1 \rb)
            + \lb( \bnabla \psi \rb)^{2} 
       \rb)\tens{1}
     - 2 \bnabla \psi \otimes \bnabla \psi 
\rb)
\;.
\label{stresstens_def}  
\end{eqnarray}

The force $\vec{F}^{\psi}_p$ acting on the $p$-th macroion is obtained by 
integrating the normal component of the stress tensor \eq{stresstens_def} 
over the particle surface $\partial G_p$, \ie
\begin{equation}
\label{pot_force_def}
\vec{F}^{\psi}_p=
\oint_{\partial G_p} \tens{T} \cdot \hat{\vec{n}}_p \, dS
\;.
\end{equation}
It may be verified by straight forward calculation
making use of the PB \eq{PBE} that
the divergence of the stress tensor vanishes.
Therefore, the integration in \eq{pot_force_def}
needs not be taken over the particle surface;
any other surface enclosing the macroion will give the same result.

\subsection{Colloid equation of motion}
\label{ssec-collmot}

Since the shape and charge distribution of
the macroions are spherically symmetric,
only translational degrees of freedom have to be considered. 
Denoting the position
of the $p$-th macroion by $\vec{R}_p$, $p=1 \ldots N$,
its equation of motion on the diffusive time scale follows from
the force balance
\begin{equation}
\label{force_balance}
0
=
\vec{F}^{\rm D}_p + \vec{F}^{\psi}_p + \vec{F}^{\rm S}_p
\;.
\end{equation}
The electrostatic force $\vec{F}^{\psi}_p$ is calculated
from the solution of the PB equation 
as described in the previous paragraph.
Dissipative and stochastic forces model the heat bath.
The dissipative forces $\vec{F}^{\rm D}_p$ are given by
\begin{equation}
\label{diss_force_def}
\vec{F}^{\rm D}_p=-\zeta\dot{\vec{R}}_p
\;,
\end{equation}
where $\zeta=6\pi\eta a$ is the Stokes friction coefficient
of a macroion with radius $a$ in the solvent of viscosity $\eta$.
The stochastic forces are related to the dissipative drag
by the fluctuation dissipation theorem
in order to ensure the correct equilibrium distribution.
We have
\begin{equation}
\label{stoch_force_def}
\vec{F}^{\rm S}_p = \sqrt{2 k_{\rm B} T \: \zeta \,} \: \vec{\xi}_p
\;,
\end{equation}
where $T$ is the solvent temperature,
$k_{\rm B}$ is the Boltzmann constant
and ${\bf{\xi}}_p$ is an uncorrelated Gaussian white noise
with zero mean and unit variance, \ie
\begin{eqnarray}
\label{noise_corr}
\big\langle
  \vec{\xi}_p(t)
\big\rangle                                       & = &
0                                                 \\
\big\langle
  \vec{\xi}_p(t)  \,\, \vec{\xi}_q^T(t^{\prime})
\big\rangle                                       & = &
\delta_{pq} \, \delta(t - t^{\prime}) \, \tens{1} \nonumber
\;.
\end{eqnarray}

\subsection{Length and Time Scales}
\label{ssec-scales}

To render the equations describing the colloid dynamics dimensionless
it remains to introduce suitable scales to measure length and time.
An obvious choice for the length scale is the macroion radius $a$.
A physically sensible estimate of the relevant time scale $\tau$
is obtained as follows.
An average force constant $k$ for the interaction between the macroions
may be defined in terms of 
the space-averaged Laplacian of the electrostatic energy as
\begin{eqnarray}
\beta k =
\frac{1}{3} \frac{1}{V} \int_{V} \bnabla^{2} \psi \, dV
\;,
\end{eqnarray}
where the volume of integration is the computational box.
Using the PB equation and the charge density of the small ions this becomes
\begin{eqnarray}
\beta k =
\frac{1}{3} \frac{\kappa^2}{2 c_s} 
\frac{1}{V} \int_{V} (n_{+}-n_{-}) \, dV
\;.
\end{eqnarray}
Due to electro-neutrality, the integral must equal $Z N$,
so that one finally finds
\begin{eqnarray}
\beta k 
=
\frac{\lambda_B Z \eta}{a^{3}}
\;.
\end{eqnarray}
From this average force constant the time scale of the motion
is obtained as $\tau = \sqrt{\zeta/k}$.

\section{Spatial discretization and PB solution algorithm}
\label{sec-pbe}

\begin{figure}[ht]
  \begin{center}
    \epsfig{file=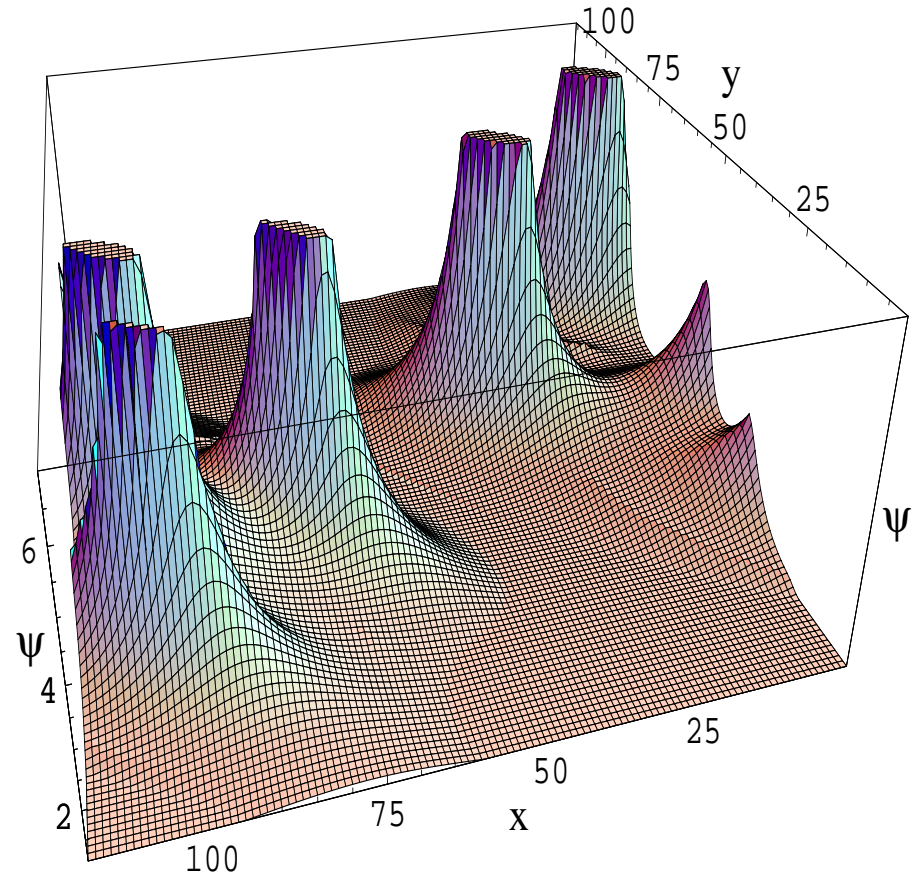, width=12cm,,height=12cm}
  \end{center}
  \caption{\label{fig_boundlay}
    Potential landscape calculated for a random configuration of
    $N=5$ macroions in a cubic box, the size of which is given here in
    units of the grid spacing. The macroions are confined to
    the plane $z=L/2$ and the potential values in this plane are shown. 
    The charge on each colloid is $Z=120$ and the screening parameter is
    $\kappa a=0.5$.
    }
\end{figure}
\begin{figure}[h]
  \begin{center}
    \epsfig{file=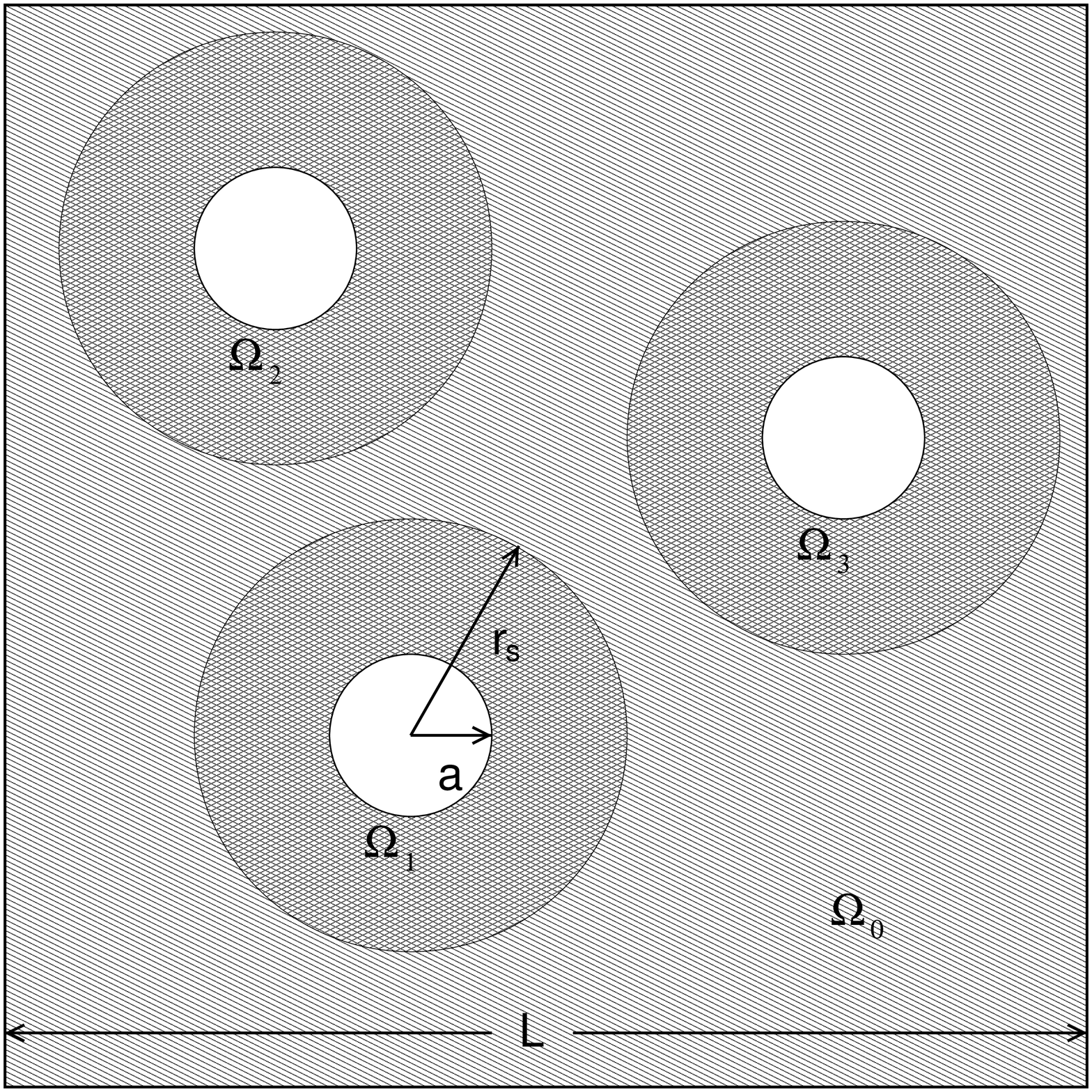, height=10cm}
  \end{center}
  \caption{\label{fig_domain}
    Sketch of the computational box, a cube of side length $L$,
    with $N=3$ macroions. 
    The domain $G$ outside the latter is partitioned into 
    spherical shells $\Omega_p$ around each macroion (crosshatched)
    and an interstitial region $\Omega_0$
    comprising the rest of the simulation box (hatched).
  }
\end{figure}

 
For large colloidal charge $Z$, 
the solution of the PB equation 
may develop rather steep boundary layers 
close to the particle surfaces 
while far away it varies only gradually
(\cf \fig{fig_boundlay}).
To provide an accurate resolution of the boundary layers
with a reasonable number of grid points, 
we follow Ref. \cite{Fushiki:JCP97-92-6700} and
define a spherical grid in a shell 
centered about each of the macroions 
overset on a Cartesian grid covering the simulation box.
The solution of the PB equation on the domain $G$ outside the macroions 
is then obtained in four steps (\cf \fig{fig_domain}):
First the PB equation is solved in each of the $N$ spherical shells 
$\Omega_{1}, \ldots, \Omega_{N}$
assuming given potential values at their outer edges.
Interpolation of these spherical shell solutions to the Cartesian grid
next yields boundary values for the interstitial region $\Omega_{0}$
which is the part of the domain $G$ 
not contained in any of the spherical shells.
With these boundary values the PB equation is then solved
in the interstitial region.
Finally, new boundary values on the outer edges of the spherical shells 
are found by interpolating back from the interstitial solution.
These steps are repeated
until a converged solution on the whole domain $G$ is obtained.
In the following we describe in turn
the discretization of the PB equation on the Cartesian and spherical grids,
the interpolation procedure between the grids, and
the method used to solve the discrete equations.

\subsection{Discretized equations on the Cartesian background grid}
\label{sec_cart}

The cubic simulation box is covered by a Cartesian background grid
with equal grid spacing $\Delta$ in the three space directions.
The size of $\Delta$ is chosen so that it matches 
the radial step size at the outer edge of the spherical shells
(\cf \sect{sec_spher}).
Typical values resulting are $\Delta = 0.2 \ldots 0.3 a$.

Denoting by $\psi_{ijk}$
the value of the potential at the grid point with indices $i,j,k$
in the $x$-, $y$-, and $z-$ directions respectively,
the PB \eq{PBE} in Cartesian coordinates
is discretized in a straight forward manner by
the central difference approximations
\begin{eqnarray}
\dpar{\psi_{ijk}}{x}                                                &\approx&
\frac{\psi_{i+1jk}-\psi_{i-1jk}}
     {2\Delta}                                                      \nonumber \\
\ddpar{\psi_{ijk}}{x}                                               &\approx&
\frac{\psi_{i+1jk}-2 \psi_{ijk}+\psi_{i-1jk}}
     {\Delta^{2}}                                                   \nonumber \\
\drpar{\psi_{ijk}}{x}{y}                                            &\approx&
\frac{\psi_{i+1j+1k}- \psi_{i+1j-1k}-\psi_{i-1j+1k}+\psi_{i-1j-1k}}
     {4\Delta^{2}} 
\label{cent-diff}  
\end{eqnarray}
and corresponding formulae for derivatives in the other coordinate directions.
The mixed derivatives do not appear in the PB equation
but are needed for the interpolation to the spherical regions
(cf. \sect{sec-interp}).
The discretized PB equation in Cartesian coordinates then becomes
\begin{eqnarray}
&\quad&
 \psi_{i+1jk}+\psi_{i-1jk}+\psi_{ij+1k}+\psi_{ij-1k}+\psi_{ijk+1}+\psi_{ijk-1}
-6\psi_{ijk}\nonumber\\
&=&
(\kappa\Delta)^{2}\sinh\lb(\psi_{ijk}\rb)
\;.
\label{PBc_disc}  
\end{eqnarray}
Values of the potential at the overlapping grid points, \ie
those points of the Cartesian grid that 
lie in one of the spherical shells $\Omega_{i}$
are obtained by interpolation from the respective spherical grid
as described in \sect{sec-interp}. 
These values are used as fixed boundary conditions 
for solving the PB \eq{PBc_disc} in the interstitial region $\Omega_{0}$. 
The second set of boundary conditions for the Cartesian problem are
the periodic boundary conditions on the surface of the computational box
which are implemented in an obvious way.

\subsection{Discretized equations on the spherical grids around the particles}
\label{sec_spher}

In a local coordinate system with the origin at the center of a macroion
the spherical shell around it is defined as the region between 
the macroion surface $r = a$ and a spherical surface at some distance $r=r_{s}$
which is taken to be the same for all of the macroions.
With the choice $r_{s} \lesssim d/2$,
an overlap of a spherical region with another particle 
is avoided for practical purposes.
To enhance the resolution near the particle surfaces, 
modified spherical coordinates are used,
where the radial coordinate $r$ is transformed 
to its inverse $w$ according to \cite{Muller:KCB26-28-257}
\begin{eqnarray}
r=\frac{1}{w}
\;,\;
dr=-\frac{dw}{w^{2}}
\;.
\label{w}  
\end{eqnarray}
To fix the step sizes in the $w$-, $\theta$-, and $\phi$-directions
we first make a choice for $\Delta_{w}$ so that
the radial grid spacing at the particle surface,
$\Delta_{r}|_{r=a} = 1 / \Delta_{w}$,
is some small fraction of the particle radius,
a typical value being $\Delta_{r}|_{r=a}  = 0.04 a$.
Then $\Delta_{\theta}$ and $\Delta_{\phi}$
are adjusted so that at the equator of the coordinate system
the grid points have approximately equal distances.
Together with the value of $r_{s}$ this also determines
the number of grid points $N_{w}$, $N_{\theta}$, and $N_{\phi}$
in the interior of the sliced spherical shell.
The potential values at the grid points here are denoted as
$\psi_{ijk} = \psi(i \Delta_{w}, j \Delta_{\theta}, k \Delta_{\phi})$.
Extra boundary points with indices $i=0,N_{w}+1$ correspond to 
the outer edge of the spherical region and the particle surface, respectively.
The points on the polar axis also require a special treatment
and, hence, are labeled by $j=0,N_{\theta}+1$.

The PB \eq{PBE} in the modified spherical coordinate system becomes
\begin{eqnarray}
 w^{2}\ddpar{\psi}{w}
+\ddpar{\psi}{\theta}+\frac{1}{\tan\!\theta}\dpar{\psi}{\theta}
+\frac{1}{\sin^{2}\!\!\theta}\ddpar{\psi}{\phi}
=\frac{\kappa^{2}}{w^{2}}\sinh\lb(\psi\rb)
\;.
\label{PBs}  
\end{eqnarray}
Using central differencing, its discretized form reads
\begin{eqnarray}
& &
  \psi_{i+1jk}
+ \psi_{i-1jk}
\nonumber \\[\medskipamount]
&+&
  \psi_{ij+1k}\Biggl[
                \lb(\frac{\Delta_{w}}{w_{i}\Delta_{\theta}}\rb)^{2}
              + \frac{\Delta_{w}^{2}}{2w_{i}^{2}\tan\theta_{j}\Delta_{\theta}}
            \Biggr]
+ \psi_{ij-1k}\Biggl[
                \lb(\frac{\Delta_{w}}{w_{i}\Delta_{\theta}}\rb)^{2}
              - \frac{\Delta_{w}^{2}}{2w_{i}^{2}\tan\theta_{j}\Delta_{\theta}}
              \Biggr]
\nonumber \\
&+&
  \psi_{ijk+1} \lb(\frac{\Delta_{w}}{w_{i}\sin\theta_{j}\Delta_{\phi}}\rb)^{2}
+ \psi_{ijk-1} \lb(\frac{\Delta_{w}}{w_{i}\sin\theta_{j}\Delta_{\phi}}\rb)^{2}
\nonumber \\
&-&
2 \psi_{ijk}\Biggl[
                1
              + \lb(\frac{\Delta_{w}}{w_{i}\Delta_{\theta}}\rb)^{2}
              + \lb(\frac{\Delta_{w}}{w_{i}\sin\theta_{j}\Delta_{\phi}}\rb)^{2}
            \Biggr]
\nonumber \\
&=&
\frac{\kappa^{2}\Delta_{w}^{2}}{w_{i}^{4}}\, \sinh\lb(\psi_{ijk}\rb)
\;.
\label{PBs_disc}  
\end{eqnarray}
On the polar axis of the spherical coordinate system,
where $\theta = 0 \;\mbox{or}\; \pi$,
the derivative with respect to $\phi$ is not defined and
its coefficient in the PB \eq{PBs} diverges.
A valid discretization at these points of the spherical grid
is obtained from the integral version of the PB equation
which by applying the Gauss statement becomes
\begin{eqnarray}
\oint_{S}
  \bnabla \psi \cdot \hat{\vec{n}} 
\, dS
= 
\kappa^{2}
\int_{V}
  \sinh\lb(\psi\rb)
\, dV
\,.
\label{PBint}
\end{eqnarray}
A discrete relation for the grid points on the polar axis,
$\psi_{ijk}$ with $j=0,N_{\theta}+1$, 
is obtained by considering a volume of integration 
bounded by the mid planes perpendicular to the grid lines 
connecting the point $(i,j,k)$ to its nearest neighbors.
For the points on the polar axis this is a frustum of pyramid 
the basis of which is formed by a regular polygon with $N_{\phi}$ edges.
Due to the smallness of the integration volume and its faces 
variations of $\psi$ in the volume and 
variations of $\bnabla \psi$ on each face may be neglected.  
Approximating $\psi$ by the value at the grid point 
and $\bnabla \psi$ by the value at the center of each face 
the following discrete expressions result:
\begin{eqnarray}
&&
  \psi_{i+10}+\psi_{i-10}
- \lb[2+\lb(\frac{2\Delta_{w}}{w_i\Delta_{\theta}}\rb)^2\rb]\psi_{i0}
+ \frac{2\Delta_{\phi}}{\pi}\lb(\frac{\Delta_{w}}{w_i\Delta_{\theta}}\rb)^2
  \sum_{k=1}^{N_{\phi}}\psi_{i1k}\nonumber\\&&
\qquad\approx\,
  \lb(\frac{\kappa\Delta_{w}}{w_i^2}\rb)^2\sinh\lb(\psi_{i0}\rb) \nonumber\\  
&&
  \psi_{i+1N_{\theta}+1}+\psi_{i-1N_{\theta}+1}
- \lb[2-\lb(\frac{2\Delta_{w}}{w_i\Delta_{\theta}}\rb)^2\rb]\psi_{iN_{\theta}+1}
+ \frac{2\Delta_{\phi}}{\pi}\lb(\frac{\Delta_{w}}{w_i\Delta_{\theta}}\rb)^2
  \sum_{k=1}^{N_{\phi}}\psi_{iN_{\theta}k}\nonumber\\&&
\qquad\approx\,
  \lb(\frac{\kappa\Delta_{w}}{w_i^2}\rb)^2\sinh\lb(\psi_{iN_{\theta}+1}\rb)
\label{PBp_disc}
\end{eqnarray}
from which the values of the potential 
at grid points on the polar axis are computed.
Since at these points the polar angle $\phi$ is not defined,
the third index $k$ has been dropped.

The boundary condition \eq{BC_coll} on the particle surface
translates into
\begin{eqnarray}
\lb.\dpar{\psi}{w}\rb|_{w=1/a}
=
Z\lambda_{B}
\;,
\label{BCs}  
\end{eqnarray}
with the discretized form
\begin{eqnarray}
\psi_{N_{w}+1jk}=\psi_{N_{w}jk}+Z\lambda_{B}\Delta_{w}
\;,
\label{BCs_disc}  
\end{eqnarray}
where a one-sided difference approximation has been employed.
The other set of boundary conditions are 
the values of the potential at the outermost surface $r = r_{s}$ 
of the spherical region
$\psi_{0jk}$, which are determined by interpolation from the Cartesian grid.

\subsection{Interpolating between the coordinate systems}
\label{sec-interp}

\begin{figure}[ht]
  \begin{center}
    \epsfig{file=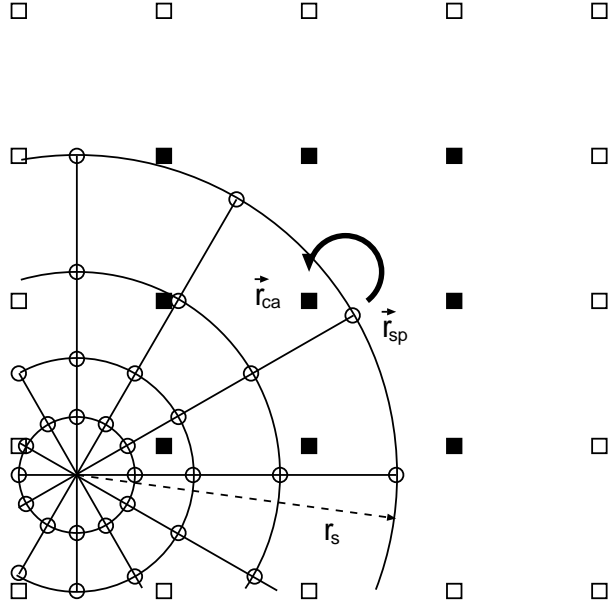, height=8cm}
  \end{center}
  \caption{\label{fig_ctos}
    At each outer boundary point $\vec{r}_{sp}$ of the spherical shells, 
    the value of the potential is defined by interpolation using 
    the closest point $\vec{r}_{ca}$on the Cartesian grid 
    (indicated by the arrow) 
    and its nearest and next-nearest neighbors (black squares).
    Note that in  three dimensions this amounts to 
    a total of 19 Cartesian grid points.
  }
\end{figure} 
After a pass on the Cartesian grid
the values of the potential at the outer spherical boundary points 
with $r=r_{s}$ 
have to be related to the values just obtained for the Cartesian grid points. 
This is done in the following manner (\cf \fig{fig_ctos}). 
For each point $\vec{r}_{sp}$ on the edge of the spherical shell,
first the closest point $\vec{r}_{ca}$ on the Cartesian grid is sought. 
Then all first and second order derivatives of the potential $\psi$ 
at that point are calculated according to \eq{cent-diff} 
using its 18 nearest and next-nearest neighbors. 
Introducing the components $q_{\alpha}, \; \alpha=x,y,z$
of the difference $\vec{r}_{sp}-\vec{r}_{ca}$, \ie
$\vec{r}_{sp}-\vec{r}_{ca}
= \sum_{\alpha=x,y,z} q_{\alpha}\hat{e}_{\alpha}$
the value at the spherical boundary point is then approximated as
\begin{eqnarray}
 \psi(\vec{r}_{sp})
=\psi(\vec{r}_{ca})
+ \sum_{\alpha}
    \dpar{\psi}{q_{\alpha}}q_{\alpha}
+ \frac{1}{2}
  \sum_{\alpha,\beta}
    \drpar{\psi}{q_{\alpha}}{q_{\beta}}q_{\alpha}q_{\beta}
\;.
\label{Taylor}  
\end{eqnarray}
These values then serve as a constant-potential boundary condition for 
the solution of the problem in the spherical shell. 

Once the PB equation in the spherical shells has been solved,  
the values at the overlapping grid points, \ie
the Cartesian grid points falling inside a spherical region,
have to be recalculated.
To avoid the use of one-sided differences 
close to the edges of the spherical regions,
this recalculation is done only for
the Cartesian points lying inside a somewhat smaller sphere
with radius $r_{a}=r_{s}-\Delta_{r}|_{r=r_{s}}/2$, 
where $\Delta_{r}|_{r=r_{s}}/2$ is 
the radial spacing at the edge of the spherical shell.
For all of these points the potential values are interpolated 
in the same way as before 
in the interpolation from the Cartesian to the spherical grid.
Note that even though points in the interior of the overlapping region
never appear in the Cartesian grid calculation,
their values may still be needed for the interpolations 
in case two or more spherical regions overlap.

While in principle it may happen that a Cartesian grid point
lies inside the particle itself,
where the spherical grid is not defined,
according to our experience this hardly ever occurs in practice
for suitably chosen parameters.
To avoid uncontrolled abortion of the program,
we nevertheless assign 
the average of all potential values on the particle surface to such points 
and monitor the frequency of their occurrence by a counter.
\begin{figure}[h]
  \begin{center}
    \epsfig{file=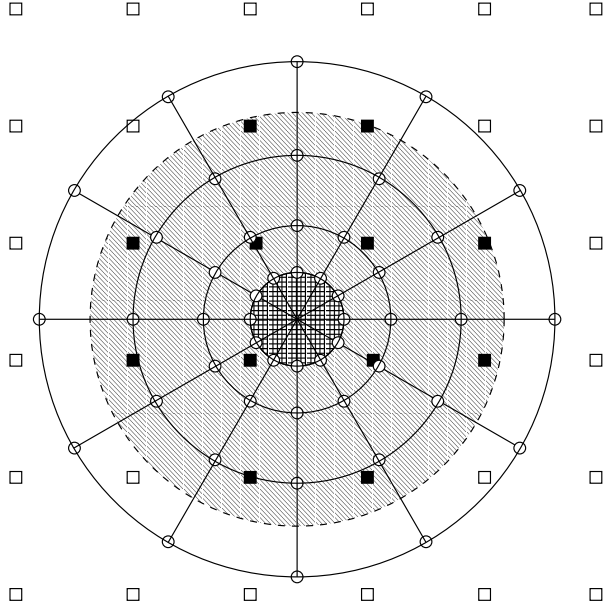, height=8cm}
  \end{center}
  \caption{\label{fig_stoc}
    At the overlapping grid points (solid squares) of the Cartesian grid,
    the values of the potential
    are defined by interpolation from the spherical grid.
  }
\end{figure} 

\subsection{Iterative Solution of the Discrete Equations}

On each of the grids the discretized PB equation is solved by
a nonlinear {\em Successive Over-Relaxation} (SOR) method \cite{Ortega:ISNE-70}.
In each step of the iteration, first the {\em residuum}
\begin{eqnarray}
\mathcal{R}_{ijk}
&=&
\;\;a\psi_{i+1jk}+b\psi_{i-1jk}+c\psi_{ij+1k}+d\psi_{ij-1k}+e\psi_{ijk+1}+f\psi_{ijk-1}+g\psi_{ijk}\nonumber\\
&&\!-h\sinh\lb(\psi_{ijk}\rb)
\label{res}  
\end{eqnarray}
is calculated at each interior grid point.
Here $a \ldots h$ denote the coefficients
in the discretized PB equation, 
\eq{PBc_disc} for the interstitial region $\Omega_{0}$ and 
\eq{PBs_disc} for the spherical shells $\Omega_{1}, \ldots, \Omega_{N}$.
New values of the potential 
are used in the calculation as soon as they are available. 
Then the solution is updated according to
\begin{eqnarray}
\psi_{ijk}(l+1) 
= 
\psi_{ijk}(l) - \omega \frac{\mathcal{R}_{ijk}}{g}
\;,
\label{overcorrect}  
\end{eqnarray}
where $l$ is the iteration count.  
Finally, the boundary points are updated
(\cf Eqs. \emb{PBp_disc} and \emb{BCs_disc})
except those where the value of the potential is fixed, of course).

For the over-relaxation parameter $\omega$ we found that 
a value of $\omega = 1.5$ was suitable for most cases, 
while sometimes it had to be reduced to $\omega = 1.3$.  
The iteration is terminated when 
the maximum of the absolute value of the residuum for all grid points
drops below some desired accuracy.  
More precisely the solution is considered converged when 
$max_{ijk} (|\mathcal{R}_{ijk} / \psi|) \le 10^{-6}$.  
Note that it is not necessary to iterate to convergence 
in each of the repeatedly solved sub-domain problems 
for $\Omega_{0}, \ldots, \Omega_{N}$. 
Only in the last pass of this repetition, 
the full convergence must be ensured. 
In fact, we usually took only a single SOR step for each sub-domain problem. 
The necessary number of repetitions of the whole procedure then was 
a few hundred times for $\mathcal{O}(100^{3})$ Cartesian grid points.

\section{Force Calculation}
\label{sec-force}

As discussed in \sect{ssec-force},
the electrostatic forces are obtained from
an integration of the normal component of the stress tensor
over an arbitrary surface enclosing the particle.
This integration is best performed 
in the spherical shell $\Omega_{p}$ around each particle
where the surface of integration is taken to be 
a sphere with radius $r_f = 1 / w_f$.
Introducing local base vectors
\begin{eqnarray}
\hat{\vec{e}}_{w}      = \pmatrix{ \sin\theta \cos\phi \cr 
                                   \sin\theta \sin\phi \cr 
                                   \cos\theta}, \;
\hat{\vec{e}}_{\theta} = \pmatrix{ \cos\theta \cos\phi \cr
                                   \cos\theta \sin\phi \cr 
                                  -\sin\theta}, \;
\hat{\vec{e}}_{\phi}   = \pmatrix{-\sin\phi \cr 
                                   \cos\phi \cr 
                                   0}
\;,
\label{spherbase}  
\end{eqnarray}
which are the same for the usual and our modified spherical coordinates,
the normal vector and surface element are simply 
$\hat{\vec{n}} = \hat{\vec{e}_{w}}$ and $dS = 1/w^2$.
Thus, according to \eq{pot_force_def}, the force becomes
\begin{eqnarray}
\vec{F}
=
\frac{1}{w^2}
\int\!\!\int \Biggl [
    T_{ww}      \hat{\vec{e}_{w}}
  + T_{w\theta} \hat{\vec{e}_{\theta}}
  + T_{w\phi}   \hat{\vec{e}_{\phi}}
\, \sin\theta \Biggr ] \, d\theta d\phi
\;.
\label{forcesph}  
\end{eqnarray}
The modified spherical components of the stress tensor
are found from \eq{stresstens_def} with 
the expression of the gradient
\begin{equation}
\bnabla \psi
= 
- w^{2}                \dpar{\psi}{w}      \hat{\vec{e}_{w}}
+ w                    \dpar{\psi}{\theta} \hat{\vec{e}_{\theta}}
+ \frac{w}{\sin\theta} \dpar{\psi}{\phi}   \hat{\vec{e}_{\phi}}
\;.
\end{equation}
The result is
\begin{eqnarray}
T_{ww}&=&
  2\kappa^{2}\lb(\cosh\psi - 1\rb) 
- w^{4}\lb(\dpar{\psi}{w}\rb)^{2}
+ w^{2}\lb(\dpar{\psi}{\theta}\rb)^{2}
+ \frac{w^{2}}{\sin^{2}\!\!\theta}\lb(\dpar{\psi}{\phi}\rb)^{2}
\;,\nonumber \\
T_{w\theta}&=&
2w^{3}\:\:\dpar{\psi}{w}\dpar{\psi}{\theta}
\;,\nonumber \\
T_{w\phi}&=&
\frac{2w^{3}}{\sin\theta}\dpar{\psi}{w}\dpar{\psi}{\phi}
\;.
\label{stresssph}  
\end{eqnarray}
A discrete approximation of Eqs. \emb{forcesph} and \emb{stresssph} 
is obtained by
using the midpoint rule for the integration and
central differences for the derivatives of the potential.
The best value for the radius of the integration surface $r_f = 1/w_f$ 
turned out to be somewhere in the middle of the spherical region.  
Close to the macroion surface, the variation of the potential is strongest 
so the solution of the PB equation is less accurate there.
Close to the outer edge of the spherical region, on the other hand,
some accuracy is lost due to the interpolation between the grids.

\section{Brownian dynamics}
\label{sec-md}

The equation of motion \eq{force_balance} is advanced in time by
a stochastic Euler algorithm \cite{Kloeden:NSSDE-92,Oettinger:SPPF-96}
\begin{eqnarray}
\vec{R}_p(t+h)                              &=& 
\vec{R}_p(t) + \frac{h}{\zeta} \vec{F}_p(t) 
\;,
\end{eqnarray}
where the time step $h$ in units of the time scale $\tau$
defined in \sect{ssec-scales} is typically $h = 0.1 \tau$.
The total force is given by
\begin{equation}
\vec{F}_p 
= 
  \vec{F}^{\psi}_p 
+ \sqrt{\frac{2 k_{\rm B} T \: \zeta}{h} \,} \: \vec{\Xi}_p
\end{equation}
where the last term is a time-discrete approximation to 
the white noise defined by \eq{noise_corr}.
The $\vec{\Xi}_p$ are random numbers with
\begin{eqnarray}
\big\langle
  \vec{\Xi}_p
\big\rangle                       & = &
0                                 \\
\big\langle                       
  \vec{\Xi}_p  \,\, \vec{\Xi}_q^T 
\big\rangle \tens{1}              & = &
\delta_{pq}                       \nonumber
\;,
\end{eqnarray}
drawn independently at each time step.
Since we are interested only in calculating averages from the trajectories,
only the first two moments of the distribution of $\vec{\Xi}_p$ 
are important \cite{Kloeden:NSSDE-92,Oettinger:SPPF-96}.
Hence, it is most convenient to generate these random numbers from
a uniform distribution on the interval $[-\sqrt{3},+\sqrt{3}]$
which has zero mean and unit variance.

\section{Validation of the method}
\label{sec-test}

\subsection{Forces Between Two Isolated Macroions}

\begin{figure}[ht]
  \begin{center}
    \epsfig{file=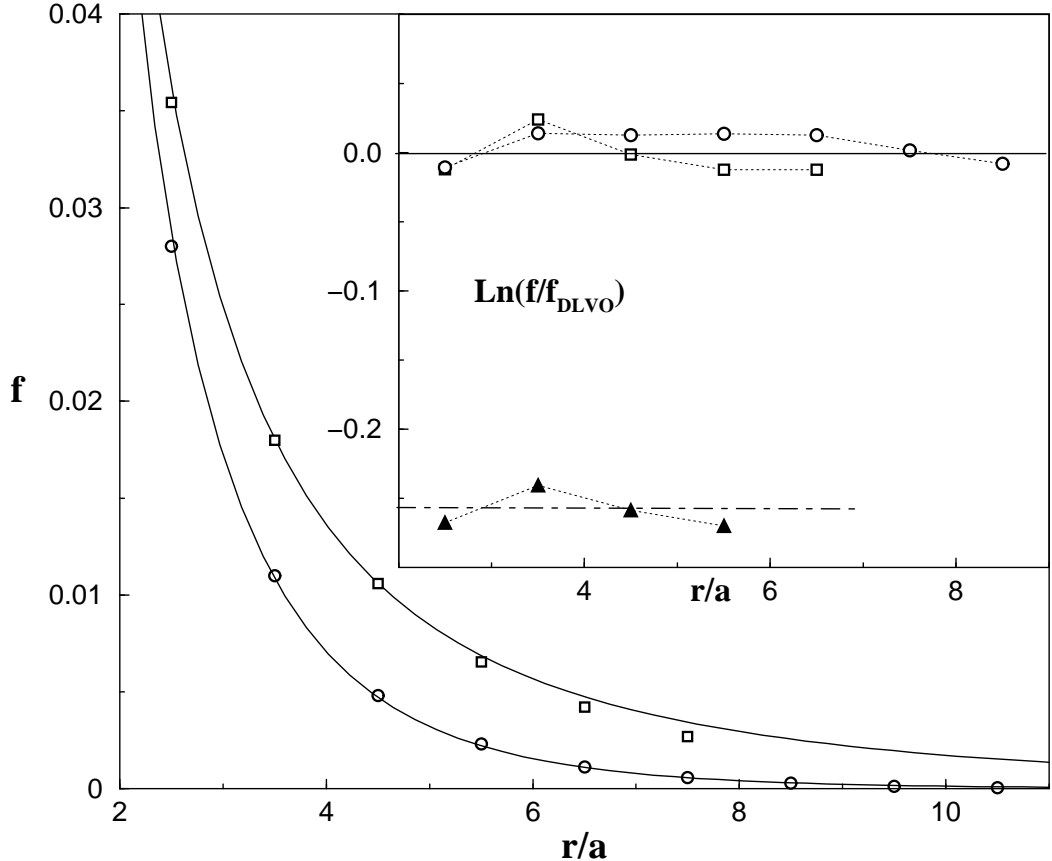, width=0.85\textwidth,angle=-90}
  \end{center}
  \caption{\label{fig_dlvoforce}
    Dimensionless force $f = \beta a F$ between two macroions 
    placed in a large simulation box 
    as function of their separation $r/a$ (symbols) compared to 
    the pair-force $f_{_{DLVO}}$ according to \protect\eq{dlvoforce} 
    predicted by linearized DLVO theory (solid lines).
    The simulation parameters are 
    $Z=4$ and $\kappa a = 0.5$ (circles),
    $Z=4$ and $\kappa a = 0.1$ (squares), and
    $Z=400$ and $\kappa a = 0.5$ (triangles).
    Inset: 
    the ratio $f/f_{_{DLVO}}$ has a constant value of 1 
    (that is $\ln(f/f_{_{DLVO}})=0$) when
    the simulation results agree with the DLVO force law, \ie for small $Z$.
    Agreement up to a constant scale factor is still found for large $Z$
    as revealed by 
    some other constant value $\ln(f/f_{_{DLVO}}) \ne 0$.
  }
\end{figure}
We first consider two isolated macroions,
for which DLVO theory \cite{Verwey:TSLC-48,Bell:JCIS33-70-335} predicts 
a repulsive double-layer interaction of Yukawa type with the potential
\begin{equation}
\label{dlvopot}
\beta U(r) 
= 
U_{0} \frac{\lambda_{B}}{r} e^{-\kappa r}
\;, \quad \mbox{where} \quad
U_{0}
=
\lb( \frac{Z e^{\kappa a}}{1+\kappa a} \rb)^{2}
\end{equation}
and the force
\begin{equation}
\label{dlvoforce}
\beta a F_{_{DLVO}}(r) 
= 
U_{0} (1 + \kappa r) \frac{\lambda_{B} a}{r^2}
e^{-\kappa r}
\;.
\end{equation}
These expressions are derived from the linearized PB equation
and hence expected to hold true if $Z \lambda_{B} / a \ll 1$.

In \fig{fig_dlvoforce} 
the force $\vec{F}$ calculated numerically from 
the solution to the full nonlinear PB equation (symbols)
is compared to the DLVO force law \eq{dlvoforce} (solid lines).  
Good agreement between both is found 
for small macroionic charge $Z=4$ (circles and squares).  
The slight deviations for the case with large screening length 
are due to the finite size of the computational box.  
For large macroionic charge $Z=400$ (triangles), however, 
there is a pronounced difference.  
By plotting the ratio $F/F_{_{DLVO}}$ it is seen that 
this difference lies entirely in the strength of the force 
and not in its dependence on the distance between the macroions.  
This suggests that even for large macroionic charge 
the force between two isolated macroions is of Yukawa form 
but with a \emph{renormalized} macroion charge $Z_{\eff}$
instead of the bare charge $Z$ (\cf \sect{sec-cell}).  
The validity of the linearization of the PB equation employed by DLVO theory 
can be checked directly by 
looking at the maximum value attained by the potential; 
the approximation $\sinh \psi \approx \psi$ is valid for $\psi \ll 1$.  
For the cases with $Z=4$ the potential does not exceed 
0.01 and 0.06, respectively for $\kappa a = 0.5$ and $\kappa a = 0.1$
so that the linearized PB equation is applicable.
For the case with $Z=400$, in contrast, the potential reaches a value of 5.5 
and therefore a linearization of the PB equation cannot be justified.

\subsection{Effective Forces}

\begin{figure}[ht]
  \begin{center}
    \epsfig{file=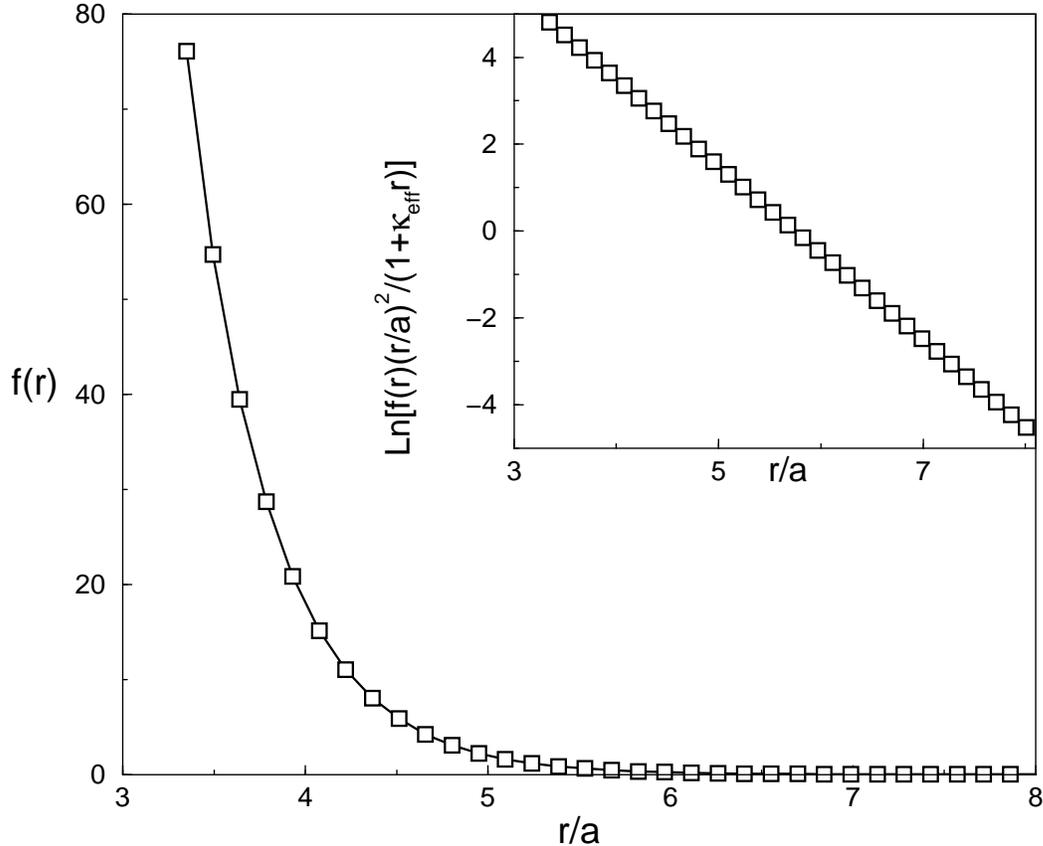, width=0.835\textwidth,angle=-90}
  \end{center}
  \caption{\label{fig_effforce}
    Dimensionless effective pair-force $f = \beta a F_{AB}$
    between two out of $N=108$ macroions 
    arranged in a FCC configuration.
    The volume fraction is $\eta=0.03$ 
    which gives a mean distance of $d_m=5.2a$.
    The bare charge of the macroions the screening parameter are 
    $Z=3000$ and $\kappa a=2.0$.
    Note that for these parameters, $F_{AB}$
    depends only on the particle separation $r/a$.
    Inset: Force multiplied by $(r/a)^{2}/(1+\kappa_{\eff} r)$ 
    and plotted logarithmically so that the Yukawa interaction appears as
    a straight line with slope $-\kappa_{\eff}$.
    This scaling clearly shows that the interaction is Yukawa-like.
    From a fit, the effective force parameters are obtained as
    $Z_{\eff}=1080$ and $\kappa_{\eff}=1.98$.
  }
\end{figure} 
For the many-particle problem of a colloidal suspension
it may be expected that the interactions 
effectively reduce to pairwise additive forces
whenever the Debye length $\kappa^{-1}$ is small compared to
the mean distance between the particles $d_m$.
The effective pair-forces are obtained
for a fixed configuration of macroions 
in the following way \cite{Belloni:private-02}:
Choosing two particles A and B, 
one first calculates the total force $\vec{F}_{B}^{1}$ 
acting on particle B with particle A present.
Then particle A is removed leaving all the other particles in place
and one again calculates the total force $\vec{F}_{B}^{0}$ 
acting on particle B now with particle A removed.
The force exerted by particle A on particle B then is just
the difference between the two and
varying the position of particle A results in the effective force curve: 
\begin{eqnarray}
\vec{F}_{AB}(\vec{r})
=
\vec{F}_{B}^{1}(\vec{r})-\vec{F}_{B}^{0}
\,,\qquad
\vec{r}
=
\vec{R}_{A} - \vec{R}_{B}
\;.
\end{eqnarray}
With this procedure, many-body interactions, if present, 
are folded into an effective pair interaction.
If the true interactions in the system are pairwise additive, the
resulting effective interaction is by construction identical to the
true pairwise interaction potential.
That is, it is independent of the direction of $\vec{r}$ 
and also independent of the arrangement of 
the surrounding particles (all particles but A and B).

For macroions arranged in a FCC configuration, the resulting effective force 
calculated by numerical solution of the PB equation
is shown in \fig{fig_effforce}.
The system parameters have been chosen to satisfy $\kappa^{-1} \ll d$
where pairwise additivity is expected.
The form of the interactions is again clearly seen 
to be of Yukawa type like in \eq{dlvoforce}
with some effective force parameters $Z_{\eff}$ and $\kappa_{\eff}$
in place of the bare charge $Z$ and the inverse Debye length $\kappa$.

To verify the pairwise additivity of the effective interaction
this calculation has been repeated for the same parameters
but taking different directions $\hat{\vec{r}} = \vec{r}/|\vec{r}|$ 
in a FCC configuration
and also with the configuration of the surrounding particles changed to BCC.
In all cases the same effective force law was found
confirming that many-body interaction are indeed absent
in this parameter regime.
When the screening parameter $\kappa a$ was reduced, however,
a very different behavior results:
The effective interaction becomes configuration dependent 
and exhibits a cut-off like feature at $\kappa^{-1} \approx d$.
Such behavior is the manifestation of 
a many-body nature of the interactions 
as discussed in detail in Refs.~\cite{Dobnikar:NN1-02,Dobnikar:NN2-02}.

\subsection{Charge Renormalization}
\label{sec-cell}

\begin{figure}[ht]
  \begin{center}
    \epsfig{file=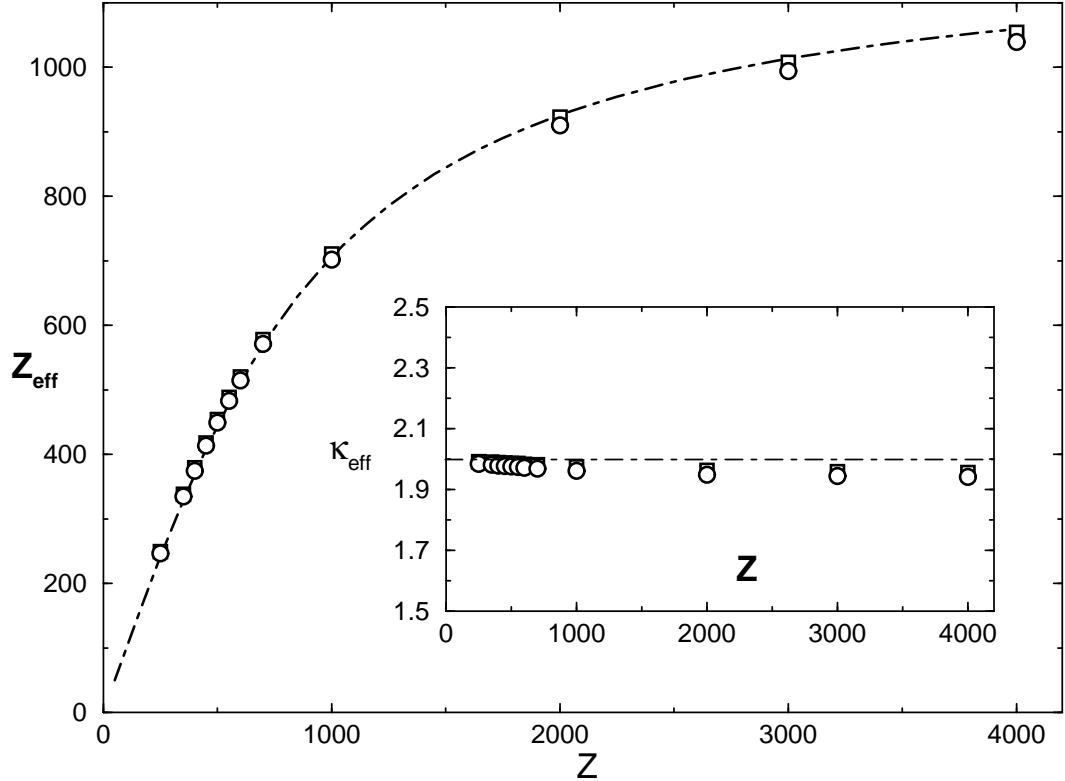, width=0.765\textwidth,angle=-90}
  \end{center}
  \caption{\label{fig_zeff}
    Effective force parameters $Z_{\eff}$ and $\kappa_{\eff}$ 
    for a Yukawa-type pair interaction 
    as functions of the bare charge $Z$
    calculated for macroions in 
    FCC and BCC configurations (circles and squares respectively).
    The simulation parameters are $N=108$ and $54$ 
    (for FCC and BCC configurations respectively),
    $\eta=0.03$, and $\kappa a=2.0$.
    There is almost perfect agreement with
    the prediction of the spherical cell model \cite{Alexander:JCP80-84-5776} 
    (dot-dashed lines).
  }
\end{figure} 
As shown in the previous section, for $\kappa^{-1} \ll d_m$
the form of the effective interaction in a colloidal crystal
is of the Yukawa type predicted by DLVO theory (\cf \eq{dlvoforce}).
Except when $Z\lambda_{B}/a$ is small, however,
the bare charge $Z$ and inverse screening length $\kappa$
have to be replaced by 
effective force parameters $Z_{\eff}$ and $\kappa_{\eff}$
to be determined by a fit to the calculated force curves.
Repeating this procedure for
different bare charges results in a charge renormalization curve 
$Z_{\eff}$ as function of $Z$ as shown in \fig{fig_zeff} (symbols).
For small bare charge, of course, $Z_{\eff} = Z$ as expected,
but for large bare charge,
$Z_{\eff}$ tends towards a constant value.
The dependence of $\kappa_{\eff}$ on the bare charge is only very weak.

\begin{figure}[ht]
  \begin{center}
    \epsfig{file=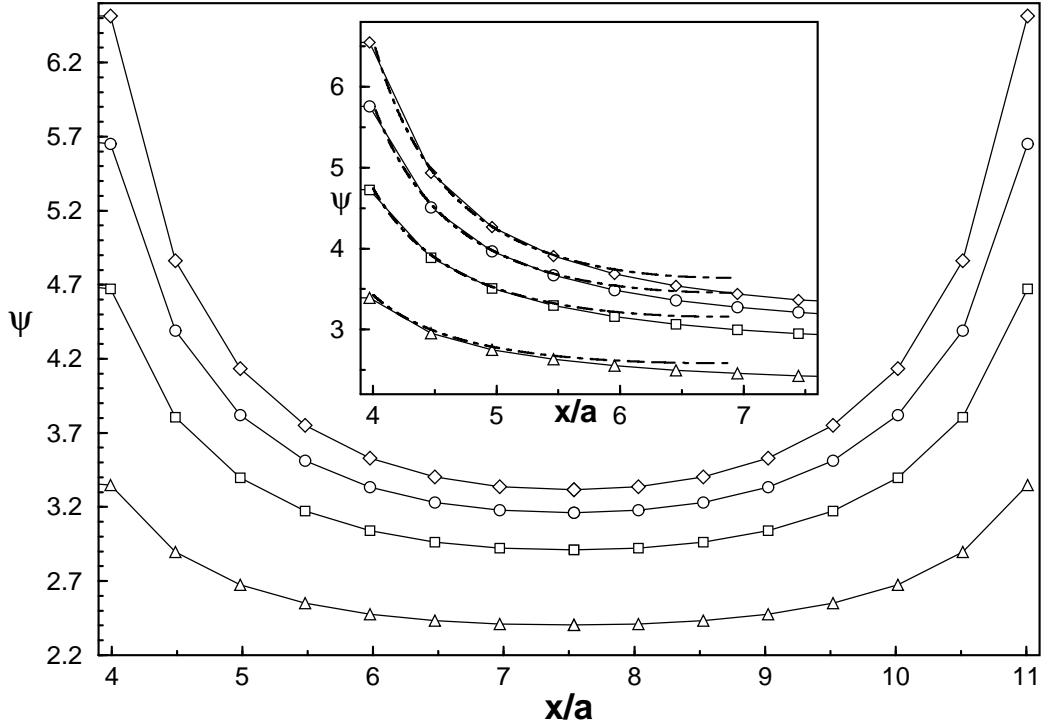, width=0.71\textwidth,angle=-90}
  \end{center}
  \caption{\label{fig_pot}
    Electrostatic potential $\psi$ 
    for $N=108$ macroions arranged in a FCC configuration 
    on a line corresponding to the 100 crystallographic direction 
    with two particles located at $x=3.0$ and $x=12.0$ 
    (symbols with solid lines to guide the eye).  
    The system parameters are: volume
    fraction $\eta=0.023$, screening parameter $\kappa a=0.3$,
    and macroion charge $Z = 21, 41, 61, 81$ from bottom to top curve.  
    Inset: comparison to the potential
    calculated from the spherical cell model (dash-dotted lines).  
    The agreement is good except near the cell edge 
    where the spherical approximation becomes invalid.
    }
\end{figure} 
The idea of charge renormalization has originally been proposed for
a cell model \cite{Alexander:JCP80-84-5776}, 
where only a single Wigner-Seitz cell of the colloidal crystal is considered.
To facilitate solution of the PB equation, 
furthermore the complicated geometry of the true Wigner-Seitz cell 
is replaced by a sphere.
The predictions of the spherical cell model for the effective parameters 
$Z_{\eff}$ and $\kappa_{\eff}$ (dash-dotted lines in \fig{fig_zeff})
agree very well with the values obtained from 
the full numerical calculation (symbols).
Again, there is no significant dependence on the crystal structure
indicating the absence of many-body effects.

A more direct comparison to the cell model can be made by 
looking at the electrostatic potential $\psi$ itself. 
In \fig{fig_pot} the numerical result for 
the potential along a line between two out of 108 macroions 
in an FCC configuration is shown for 
several values of the bare charge $Z$ (symbols).
The agreement to the spherical cell model result (dash-dotted line) 
is very good, except near the cell edge, 
where the spherical approximation leads to 
a higher value of the potential in the cell model.
This is due to the fact that 
the half-width of the actual FCC Wigner-Seitz cell in the direction shown
is $b=4.5$ while
the size of the spherical cell with equal volume is only $R=3.5$.

\subsection{Melting Point}

\begin{figure}[ht]
  \begin{center}
    \epsfig{file=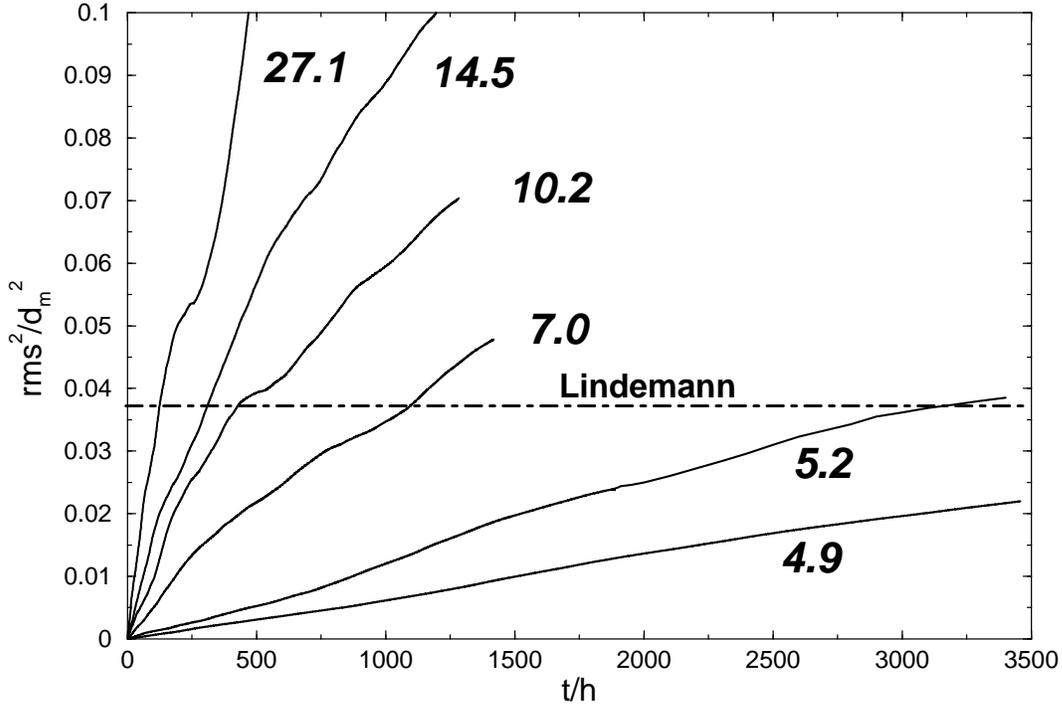, width=0.695\textwidth,angle=-90}
  \end{center}
  \caption{\label{fig_rms}
    Mean square displacement ($rms^{2}$) of $N=108$ macroions 
    as a function of time starting from an FCC configuration.  
    Simulation parameters are: volume fraction $\eta=0.03$,  
    screening parameter $\kappa a=1.0$, and 
    charges between $Z=100$ and $Z=4000$.  
    The value of the scaled temperature $10^{3} \tilde{T}$
    as defined in \protect\eq{Ttilde} is shown next to each curve.  
    According to the Lindemann criterion, melting occurs
    whenever the root mean square displacement exceeds 19\% of
    the mean distance between the macroions \cite{Robbins:JCP88-88-3286}
    (dashed line).
    }
\end{figure} 

Finally, we take a look at
a macroscopic property of the suspension
obtained from a full dynamical calculation,
namely the melting point.
For systems of point-like Yukawa particles this quantity is known from
the classic work of Robbins, Kremer, and Grest \cite{Robbins:JCP88-88-3286}.
In this work, the melting line has been calculated 
from a molecular dynamics simulation 
by using the Lindeman criterion stating that  
melting occurs whenever the root mean square displacement exceeds 
the value of 19\% of the mean distance between macroions.
The result is
\begin{equation}
\label{yukawamelt} 
\tilde{T}^{M}(\lambda) = 0.00246 + 0.000274\lambda
\end{equation}
in terms of the parameters 
\begin{eqnarray}
\lambda 
&=& 
\kappa_{\eff} d_m \qquad \mbox{and} \\
\tilde{T}(\lambda) 
&=& 
\lb( \beta U_{0} \frac{\lambda_{B}}{d_m} 
     \frac{2 \lambda^{2}\theta(\lambda)}{3}
\rb)^{-1}
\;.
\label{Ttilde}
\end{eqnarray}
The former is simply the inverse effective screening length scaled with 
the mean particle distance.
The latter represents a scaled temperature 
($kT$ in units of the Einstein phonon energy)
which is related to the strength of the interaction $U_{0}$ 
defined in \eq{dlvopot}
and, hence, the effective charge $Z_{\eff}$.
The function $\theta(\lambda)$ depends on 
the lattice-type of the crystalline phase
and is given in Tab.~I of Ref.~\cite{Robbins:JCP88-88-3286}.

In \fig{fig_rms} we show 
the mean square displacement, $rms^{2}$, of the macroions
as a function of time
as calculated from the combined PB--Brownian dynamics method.
By using the effective force parameters 
$Z_{\eff}$ and $\kappa_{\eff}$ from the previous section,
the values of $\lambda$ and $\tilde{T}$ can be calculated
so that a comparison to \eq{yukawamelt} becomes possible.
The value of $\lambda=8.1$ is practically independent of the bare charge $Z$.
The effective temperature, in contrast, depends strongly on $Z$
with values shown next to each curve in \fig{fig_rms}.
The Lindemann value $rms/d_m = 19\%$ is reached for 
$\tilde{T} \ge 5.2\cdot 10^{-3}$,
so that by the same criterion as in Ref.~\cite{Robbins:JCP88-88-3286}
the colloidal crystal melts around
$\tilde{T} = 5.1 \cdot 10^{-3} \pm10\%$. 
The melting occurs quite close to 
the saturated value of the effective charge. 
The corresponding effective temperature $\tilde{T}\approx 4.5\cdot 10^{-3}$
cannot be exceeded for our system.
We conclude that our estimate of the melting point
is in reasonable agreement with the value $5.3\cdot 10^{-3}$ 
obtained by Robbins et al. in Ref.~\cite{Robbins:JCP88-88-3286}
for Yukawa systems.

\section{Discussion}
\label{sec-conc}

The physics of charge stabilized colloidal suspensions 
poses challenging problems resulting from 
the presence of many coupled degrees of freedom.
A treatment has been possible so far only 
in limited regions of parameter space
either due to the necessity of approximations in analytical treatments
or because of practical limitations in numerical simulations.
We have here described in detail a method 
which makes accessible the regime of
large macroionic charge, low salt concentration, and
relatively high colloid volume fraction,
where many-body interactions between the macroions become important
\cite{Dobnikar:NN1-02,Dobnikar:NN2-02}.
This has been achieved by combining 
a Poisson-Boltzmann field description of the small ions
with a Brownian dynamics simulation of the charged colloidal particles.
By describing the small ions in terms of a continuous density,
the computational effort becomes independent of their number.
In this way, the main limitation, which precludes the application
of primitive model simulations to the above parameter regime
is circumvented.
At the same time, in contrast to
analytical treatments like linearized DLVO theory or cell models,
macroionic many-body effects are fully accounted for in our approach.

In the present work, we have focused on 
the case where the effective interactions are
still of a pairwise additive Yukawa form.
By comparing our simulation results to
linearized DLVO theory and cell model calculations,
which are applicable in this regime,
the method has been thoroughly validated.
Furthermore, its potential applications have been illustrated 
which reach beyond simple structural calculations 
\cite{Fushiki:JCP97-92-6700,Lowen:JCP98-93-3275}
and also include the evaluation of thermodynamic properties, \eg
the phase behavior of charged colloidal suspensions.
These applications are pursued in greater detail elsewhere 
\cite{Dobnikar:NN4-02}.

Directions for further development from a physical viewpoint are twofold.
First, on the Poisson-Boltzmann level of description 
correlations between the small ions are neglected, 
which limits the applicability to monovalent small ions and 
small coupling $\lambda_{B} \ll a$. 
The range of applicability could be extended by 
using more elaborate density functional theory 
similar to Refs.~\cite{Lowen:JCP98-93-3275,Lowen:EL23-93-673}.
Secondly, the calculation of rheological properties necessitates
inclusion of hydrodynamic effects.
The framework of a field description of the solvent 
on which our approach is built,
readily accommodates such an extension as well \cite{Shyy:CTCTP-97}.

Desirable improvements of the computational efficiency
include parallelization of the method
and the implementation of more sophisticated solvers for the discrete equations.
Several possibilities for the latter have been compared 
in Ref.~\cite{Holst:JCoC16-95-337} where 
a combination of inexact Newton and multigrid methods 
was found to be the most efficient.
A parallelization strategy that naturally fits together with 
the use of the overset spherical grids is to use
a domain decomposition approach \cite{Smith:DD-96} 
also for the Cartesian background grid.
Developments along these lines are currently pursued.


\bibliographystyle{elsart-num}
\bibliography{preamble,elyte,colloid,poly,softmat,compphys,mathmeth,misc}

\end{document}